\documentclass[epj]{svjour}
\usepackage{graphicx,epsfig}
\begin{document}


\title{Effect of B-site Dopants on Magnetic and Transport Properties of LaSrCoRuO$_6$}
\author{P S Ramu Murthy\inst{1}\and K R Priolkar\inst{1}\mail{krp@unigoa.ac.in} \and P. A. Bhobe\inst{2} \and A. Das\inst{3} \and P. R. Sarode\inst{1} \and A. K. Nigam\inst{2}}
\institute{Department of Physics, Goa University, Taleigao Plateau, Goa 403 206, India \and Tata Institute of Fundamental Research, Homi Bhabha Road, Colaba, Mumbai 400 005 India \and Solid State Physics Division, Bhabha Atomic Research Centre, Trombay, Mumbai 400 085 India}

\date{\today}

\abstract{Effect of Co, Ru and Cu substitution at B and B' sites on the magnetic and transport properties of LaSrCoRuO$_6$
have been investigated. All the doped compositions crystallize in the monoclinic structure in the space group
$P2_1/n$ indicating a double perovskite structure. While the magnetization and conductivity increase in Co and Ru
doped compounds, antiferromagnetism is seen to strengthen in the Cu doped samples. These results are explained on
the basis of a competition between linear Co-O-Ru-O-Co and perpendicular Co-O-O-Co antiferromagnetic interactions
and due to formation of Ru-O-Ru ferromagnetic networks.}

\PACS{{75.50.Ee}{Antiferromagnetic Materials} \and {75.50.Pp}{Magnetic semiconductors} \and {72.80.Ga}{Transition metal compounds} }  

\titlerunning{Effect of B-site Dopants on Magnetic and Transport Properties of LaSrCoRuO$_6$}
\authorrunning{P. S. R. Murthy et al.}
\maketitle

\section{Introduction}
Perovskites containing 3d (B) and 4d or 5d (B') transition metal cations attract attention because of the strong
competition between antiferromagnetic and ferromagnetic coupling and a complex interplay of spin charge and
orbital degrees of freedom \cite{dd}. Some of these perovskites like Sr$_2$FeMoO$_6$ and Sr$_2$FeReO$_6$ display
itinerant ferrimagnetism and large low field magnetoresistance \cite{koba,jg}. These properties can be
effectively altered by substituting the B or B' site ions with other transition metal ions \cite{sriti,feng}.

Ruthenium based double perovskites are interesting and have been studied extensively for their unusual magnetic
and transport properties. Ru(V) containing perovskites exhibit many interesting properties like itinerant
electron magnetism and co-existence of antiferromagnetic order and superconductivity
\cite{battle,kame,doi,wu,park}. Similarly Ru(IV) based double perovskites containing a transition metal ion on
the B' site present a wide variety of ground states that depend on Ru $4d$ and O $2p$ covalent mixing and the
itinerancy of $\pi^*$ electrons \cite{dass}.

LaSrCoRuO$_6$ is another double perovskite  that provides a unique opportunity to examine the interplay of
cationic order, charge balance and complex magnetic interactions between two transition metal ions
\cite{bos,mam}. Its crystal structure is composed of corner-shared CoO$_6$ and RuO$_6$ octahedra arranged in a
pseudocubic array in the rocksalt arrangement. The degree of ordering is known to affect the magnetic and
transport properties due to changes in magnetic interactions and in cationic valence. Effects of anti-site
disorder on the magnetic and transport properties due to La or Sr doping in LaSrCoRuO$_6$ have been investigated
\cite{bos,tomes,dlouha}. LaSrCoRuO$_6$ compound is a semiconductor with ideal valence states HS Co$^{2+}$ (3d$^7$
high-spin configuration) and Ru$^{5+}$ (4d$^3$) \cite{kim}. The change of composition (La or Sr doping)
introduces mobile electrons in La richer samples or holes in Sr richer samples. Magnetically these compounds are
reported to be antiferromagnetic with two magnetic face centered cubic (fcc) sublattices consisting of Co and Ru.
Both the sublattices order with type II antiferromagnetic structure which would mean that the spins in [111]
planes in succession Co-Ru-Co-Ru alternate as +/+/-/ -. This marginalizes the Co-O-Ru nearest neighbor
interactions and the ordering is governed by a competition between linear Co-O-Ru-O-Co and 90$^\circ$ Co-O-O-Co
antiferromagnetic exchange paths \cite{bos,dlouha}.

Effect of substitutions at the B or B' sites in LaSrCoRuO$_6$ have not been studied so far. In particular, the effect of antisite disorder will be very important as it will alter the magnetic interactions present in LaSrCoRuO$_6$ and perhaps result in more complex magnetic ground state. Antisite disorder in ferromagentic double perovskites like Sr$_2$FeMoO$_6$ is known to affect magnetic and transport properties of these compounds \cite{serrate}. Herein we report, structural, transport and magnetic properties of LaSrCo$_{1-x}$Ru$_{1+x}$ for $-0.3 \le x \le 0.4$ and LaSrCo$_{1-y}$Cu$_y$RuO$_6$ $y \le 0.2$ studied using X-ray diffraction (XRD), Neutron diffraction (ND), resistivity and magnetization as a function of temperature and magnetic field. The compounds studied herein have a fixed A-site variance though varying Goldschmidt tolerance factor $t $ \cite{gold} and redox active Co$^{2+/3+}$ and Ru$^{4+/5+}$ couples resulting essentially due to antisite disorder. This study elucidates the complex role of coexisting localized electrons belonging to both Co and Ru $d$ orbitals as well as some itinerant $\pi^*$ electrons of Ru:t$_{2g}$ parentage that arise due to presence of Ru$^{4+/5+}$ and Co$^{2+/3+}$ redox couples on the structural, magnetic and transport properties of substituted LaSrCoRuO$_6$. 

\section{Experimental}
\sloppy Polycrystalline samples of LaSrCo$_{1-x}$Ru$_{1+x}$O$_6$; $-0.3 \le x \le 0.4$  and LaSrCo$_{1-y}$Cu$_y$RuO$_6$; $y$ = 0.1 and 0.2  were synthesized by firing well mixed stoichiometric amounts of La$_2$O$_3$, SrCO$_3$, Co(NO)$_3$)$_2$.6H$_2$0, RuO$_2$ and CuO. These fired mixtures were then thoroughly ground and fired 1300$^\circ$C followed by slow cooling to room temperature. The process grinding and firing was repeated three times. Prior to firing, the powders of La$_2$O$_3$, RuO$_2$ and SrCO$_3$ were pre-heated at 900$^\circ$C and 700$^\circ$C respectively to get rid of any adsorbed water. The samples were characterized by XRD for their structure and phase purity. The diffracted intensities were measured in the 2$\theta$ range of 15$^\circ$ to 80$^\circ$ using Cu K$_\alpha$ radiation. The diffraction patterns were Rietveld refined using
FULLPROF suite to obtain lattice parameters. ND measurements on some of the samples were carried out on the Powder diffractometer at Dhruva Reactor, Trombay using neutrons of wavelength 1.24\AA~ in the temperature range  20K to 300K. DC magnetization was measured, both, as a function of temperature and magnetic field using the Quantum Design SQUID magnetometer (MPMS-5S). M(T) was measured in an applied field of 50 Oe and 1000 Oe in the temperature range of 5 to 300 K. The sample was initially cooled from 300K to 5 K in zero applied field and the data was recorded while warming up to 300 K in the applied magnetic field (referred to as ZFC curve) and subsequent cooling (referred to as FC curve) back to 5 K. Magnetization as a function of field was measured under sweep magnetic fields up to $\pm 5 T$ at various temperatures. Before each M(H) was recorded, the sample was warmed to 300 K and cooled back to the desired temperature. The temperature dependence of the electrical
resistivity in the range 10K - 300K was measured using a standard four probe set up.

\section{Results}

\subsection{LaSrCo$_{1-x}$Ru$_{1+x}$O$_6$}
\begin{figure*}
\centering
\includegraphics[scale=0.75]{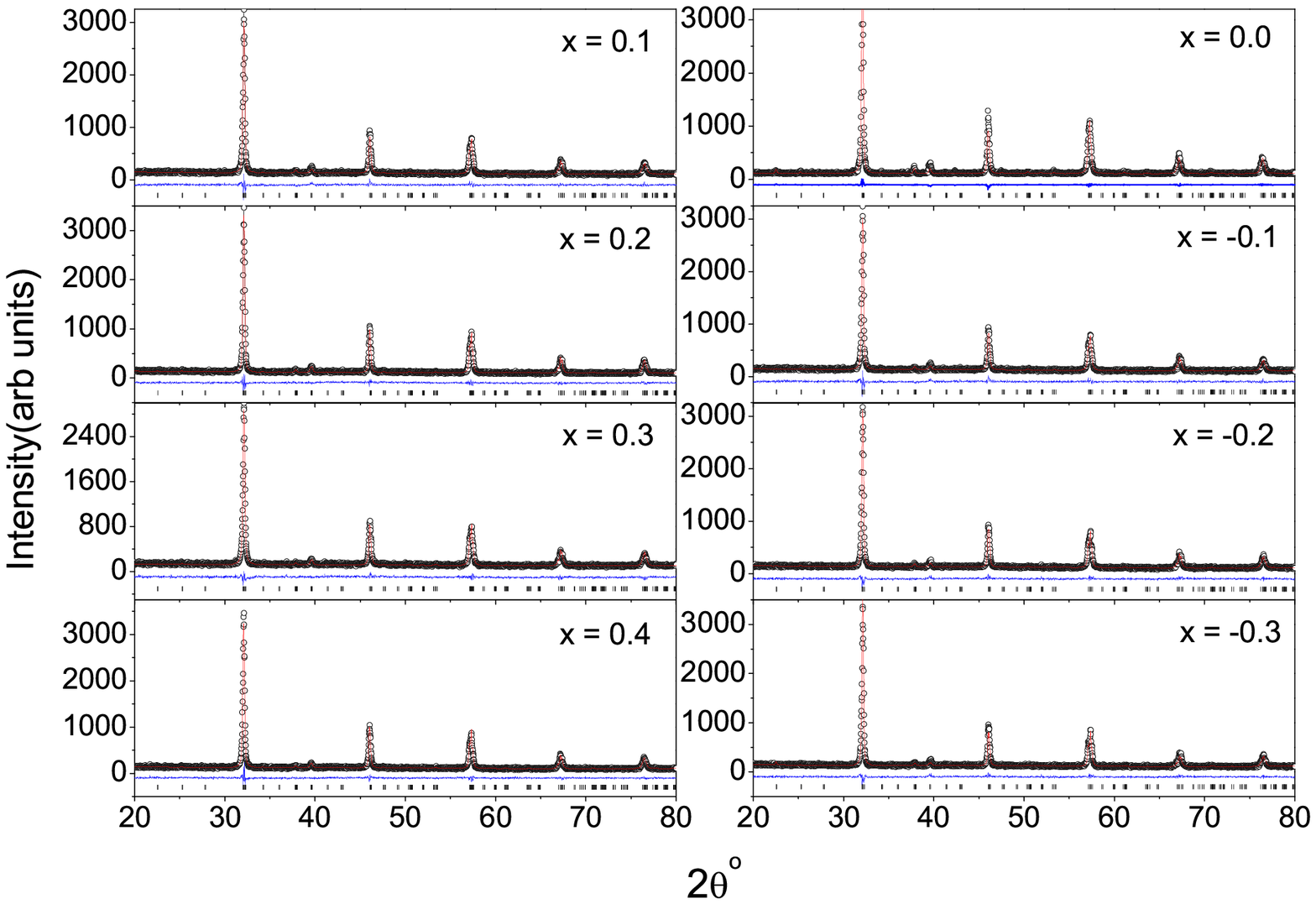}
\caption{\label{XRD}Le bail fitted XRD patterns for LaSrCo$_{1-x}$Ru$_{1+x}$O$_6$. The data points are indicated
as circles and the solid line through them is the fit to the data. The difference pattern is shown at the bottom
along with tick marks indicating Bragg reflections. }
\end{figure*}

\begin{figure*}
\centering
\includegraphics[scale=0.5]{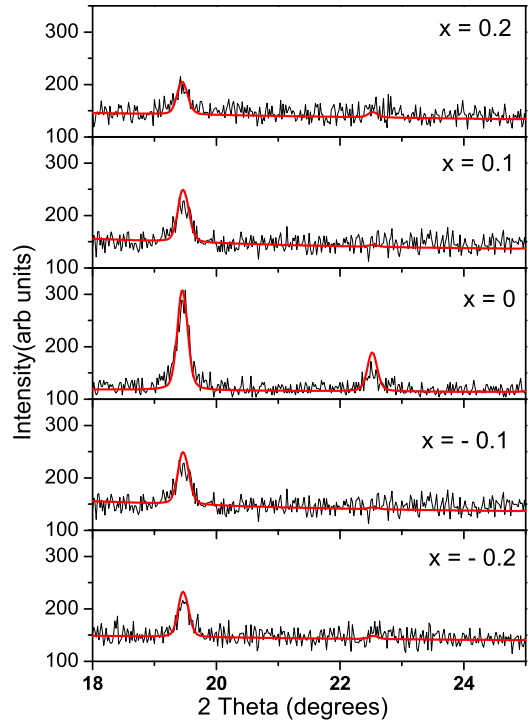}
\caption{\label{super}Variation of intensity of superlattice (${1\over2},{1\over2},{1\over2}$) reflection with increasing Ru ($x > 0$) and Co ($x < 0$) doping.}
\end{figure*}

\begin{figure*}
\centering
\includegraphics[scale=0.5]{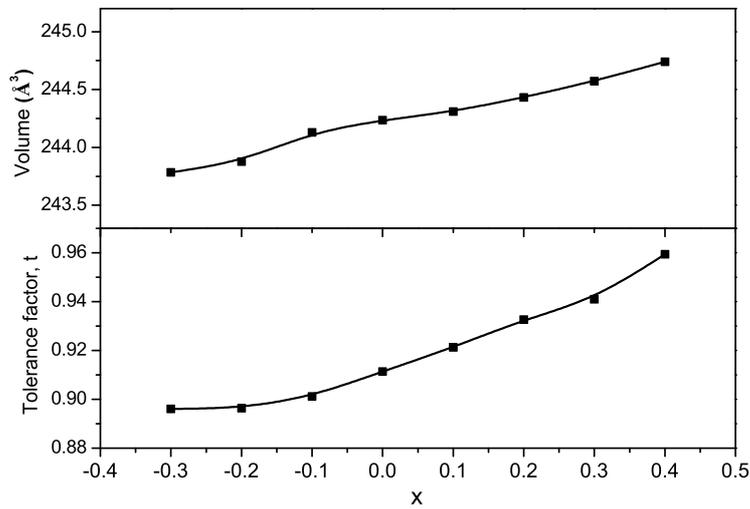}
\caption{\label{Volume}(a) Composition dependance of the unit cell volume in LaSrCo$_{1-x}$Ru$_{1+x}$O$_6$. (b)
Variation of tolerance factor $t$ with $x$ in LaSrCo$_{1-x}$Ru$_{1+x}$O$_6$.}
\end{figure*}

The stability of the ordered double perovskite phase with P2$_{1}$/n symmetry is determined within the
composition region $x$ = 0.1 to 0.4 for Ru and $x$ = -0.1 to -0.3 for Co from the Le Bail refinement of XRD
patterns which are presented in Fig. \ref{XRD}. The lattice parameters obtained from refinement are presented in
Table \ref{XRDM}. LaSrCoRuO$_{6}$ which has Co$^{2+}$ and Ru$^{5+}$ ions occupying the B-site represents a
perfectly ordered perovskite with NaCl type ordering of its B-sites cations. Substitution of Ru for Co and Co for
Ru creates not only an antisite disorder in the Co-O-Ru matrix but also changes the formal valence of the B-site ions.  
While the effect of disorder can be clearly seen in Fig. \ref{super} on the superlattice reflection (${1\over2},{1\over2} ,{1\over2}$) at $\sim$ 19.5$^\circ$ and the Co and Ru occupancies at (${1\over2},0,0$) and (${1\over2},0,{1\over2}$) sites in Table \ref{XRDM}, the linear variation of the cell volume with $x$ as seen in Fig. \ref{Volume}(a) indicates a change in the formal valence of B-site cations. It may be
noted here that Co$^{3+}$ has a smaller ionic radius than Co$^{2+}$ while the ionic radius of Ru$^{4+}$ is larger
than Ru$^{5+}$ and hence the cell volume decreases for Co doped compounds and increases with Ru doping. The
linear variation of the unit cell volume also confirms the existence of single-phase solid solutions for all
values of composition $x$ reported here. An interesting aspect of substitution of Co at the Ru site is the
comparatively large change in the monoclinic cell angle, $\beta$. It decreases from $90.05^\circ$ in the undoped
sample to $89.9^\circ$ for the Co doped sample ($x$ = -0.2)  whereas remains nearly constant for the Ru doped
samples. This could be due to the large charge and size difference between Co and Ru and/or due to a deviation
from pseudo-cubic symmetry arising because of clustering of similar B-site ions due to charge balance. With
increasing Co substitution from $x$ = -0.1 to -0.3, the electrons are likely to be removed from the Co ions
leading to a valency increase of Co from 2+ to 3+ as the Ru ions are in highly oxidized state. But this does not
satisfy the charge balance and will require some of the Ru$^{5+}$ to reduce to Ru$^{4+}$. Consequently, the
charge and the size difference between Ru and Co ion may lead to a deviation in $\beta$. This can be also be
visualized in Fig. \ref{Volume}(b) where $t$ is plotted as a function of composition $x$.
While $t$ increases from about 0.91 for $x$ = 0 to 0.95 for $x$ = 0.4 (Ru-rich compositions), it decreases to
0.89 for $x$ = -0.3 (Co-rich compositions) indicating that the substitution of Ru for Co allows for a better
matching of A-O and B(B')-O bond distances than the replacement of Ru by Co.

\begin{table*}
\caption{\label{XRDM} Room temperature structural parameters obtained from Le Bail fitting and high temperature
magnetic and transport properties for LaSrCo$_{1-x}$Ru$_{1+x}$O$_6$. Here $a$, $b$, $c$ and $\beta$ denote unit
cell parameters, N$_{\rm Ru({1\over2},0,0)}$ denotes the occupancy of Ru at (${1\over2},0,0$) site, $\mu_{eff}$ and $\Theta_{CW}$ represent effective magnetic moment and Curie Weiss temperature
respectively and $T_p$ is temperature of peak magnetization seen in ZFC curve. Numbers in parentheses are
uncertainty in the last digit.} 
\centering
\begin{tabular}{lllllllll}

\hline\noalign{\smallskip}

$x$ & -0.3 & -0.2  & -0.1 & 0 & 0.1 & 0.2 & 0.3 & 0.4\\
\noalign{\smallskip}\hline\noalign{\smallskip}
$a$ (\AA) & 5.5866(4) & 5.5861(3) & 5.5858(3) & 5.5847(4) & 5.5847(4) & 5.5846(3) & 5.5843(4) & 5.584(3)\\
$b$ (\AA) & 5.5551(6) & 5.5553(5) & 5.5555(5) & 5.5592(6) & 5.5541(5) & 5.5533(6) & 5.5524(4) & 5.5515(5)\\
$c$ (\AA) & 7.8599(5) & 7.8612(5) & 7.8634(5) & 7.8673(4) & 7.8646(4) & 7.8641(5) & 7.8629(5) & 7.8636(4)\\
$\beta$ & 89.92(2) & 89.98(1) & 89.96(2) & 90.05(2) & 90.04(2) & 90.04(1) & 90.04(1) & 90.05(1)\\
Volume (\AA$^3$) & 243.68(4) & 243.87(3) & 244.12(3) & 244.25(4) & 244.31 (3) & 244.43(3) & 244.57(4) & 244.74(3)\\
N$_{\rm Ru({1\over2},0,0)}$& 0.55 & 0.67 & 0.83 & 0.98 & 0.81 & 0.70 & 0.80 & 0.86\\
N$_{\rm Co({1\over2},0,0)}$& 0.45 & 0.33 & 0.17 & 0.02 & 0.19 & 0.30 & 0.20 & 0.14\\
N$_{\rm Co({1\over2},0,{1\over2})}$& 0.85 & 0.87 & 0.93 & 0.98 & 0.71 & 0.70 & 0.50 & 0.46\\  
N$_{\rm Ru({1\over2},0,{1\over2})}$& 0.15 & 0.13 & 0.07 & 0.02 & 0.29 & 0.30 & 0.50 & 0.54\\
$\mu_{eff}$ ($\mu_B$/fu) & 5.52(2) & 5.35(1) & -- & 5.47(2) & 5.50(2) & 5.42(2) & 5.39(2) & 5.58(3)\\
$\Theta_{CW} (K)$ & 24(1) & 37(1) & -- & -49(1) & 21(1) & 25(1) & 25(1) & 0(1)\\
$T_p$ (K) & 42(1) & 47(1) & -- & 70(1) & 53(1) & 52(1) & 54(1) & 49(1)\\
\noalign{\smallskip}\hline
\end{tabular}
\end{table*}

\begin{figure*}
\centering
\includegraphics[scale=0.75]{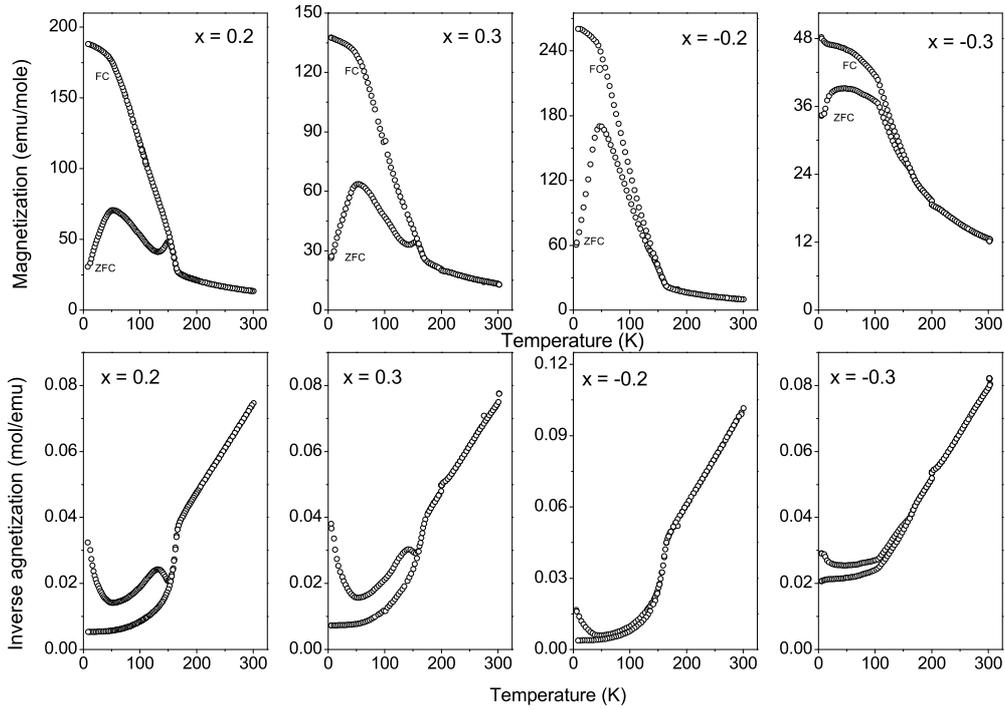}
\caption{\label{Mag} Magnetization curves for LaSrCo$_{1-x}$Ru$_{1+x}$O$_6$ during the ZFC and FC cycle recorded
at 1000 Oe.}
\end{figure*}

\begin{figure*}
\centering
\includegraphics[scale=0.5]{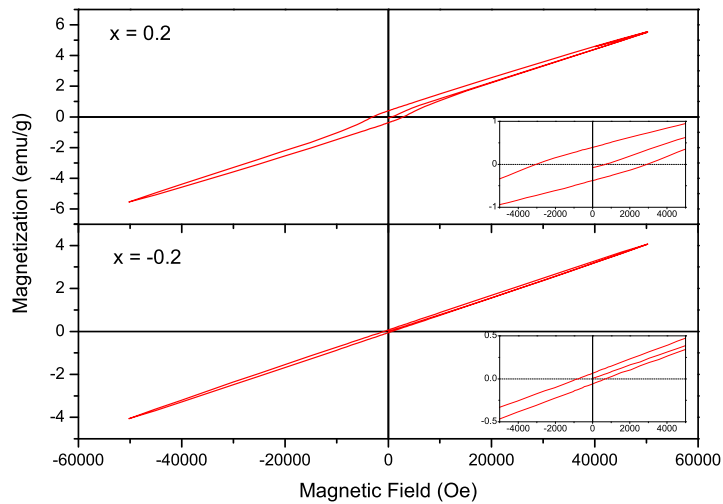}
\caption{\label{hys} Isothermal magnetization  recorded at 5K in the field range of $\pm$5T for two different
values of $x$ belonging to LaSrCo$_{1-x}$Ru$_{1+x}$O$_6$. Insets show the expanded ($\pm$4500Oe) loops.}
\end{figure*}

The structural changes occurring due to Ru and Co doping will have an effect on the magnetic properties. Fig.\ref{Mag} presents the temperature variation of magnetization measured between 5K and 300K in an applied field of 1000Oe and the magnetic susceptibility ($\chi = M/H$) calculated from thereon for four representative compounds. The difference between ZFC and FC magnetization curve reveals a complex magnetic structure. In the high temperature region (200K $\le$ T $\le$ 300K), $\chi$ was fitted to the Curie-Weiss eqaution $$ \chi = C/(T - \Theta_{CW}) $$ where $C = N_A (\mu_{eff}\mu_B)^2 )/3k_B$ is the Curie constant and $\Theta_{CW}$ is the Curie-Weiss temperature. Effective moment per formula unit ($\mu_{eff} = g[S(S+1)]^{1/2}$) calculated from $C$ and $\Theta_{CW}$ obtained for fitting for all the compounds along with undoped compound, LaSrCoRuO$_6$ and are enlisted in Table \ref{XRDM}. From the analysis of the high temperature magnetization in the doped systems, it can be inferred that Co ions are in high spin (HS) or intermediate spin (IS) state and that the magnetic interaction for all $x$ are dominated by nearest neighbor interactions of the superexchange type between HS/IS Co$^{2+,3+}$ and low spin (LS) Ru$^{4+,5+}$ ions. A second observation is that the effective magnetic moment does not vary with $x$. The nearly constant value of $\mu_{eff}$ for $x \ \in$ (-0.3 to 0.4) region can only be explained on the basis of presence
of HS and IS states for Co$^{3+}$ and Ru$^{4+}$. In the case of undoped sample, the type of long range magnetic
order is characterized by $\Theta_{CW} \sim$ -50K and a clear antiferromagnetic transition at T$_{N} \approx$ 80
K \cite{bos}. In case of Ru and Co doped samples, $\Theta_{CW}$ is positive indicating presence of additional
ferromagnetic interactions. In Ru doped samples, positive $\Theta_{CW}$ can be understood to be due to
ferromagnetic Ru$^{4+}$-O-Ru$^{4+}$ interactions. The presence of such interactions is supported by the sudden
rise of magnetization at about 160K which is close to ferromagnetic ordering temperature of SrRuO$_3$. The
presence of any SrRuO$_3$ impurity phase however, can be ruled out as no additional reflections are seen in the
diffraction pattern even at 40\% Ru doping. The competition between the ferromagnetic and antiferromagnetic
interactions can result in magnetic frustration below the peak temperature T$_{p} \sim$ 50K seen in ZFC
magnetization curve. The nature of magnetic order will be confirmed later using neutron diffraction. Positive
values of $\Theta_{CW}$ even in Co doped compounds is puzzling. It can only be understood to be due to presence
of Co$^{2+/3+}$ and Ru$^{4+/5+}$ redox pairs for charge balance. The nearly constant values of $\mu_{eff}$
supports this conjuncture well. Sudden rise seen in magnetization at around 160K seen in 20\% and 30\% Co doped
samples is also in agreement with the above view. The presence of Ru$^{4+}$ leads to the ferromagnetic Ru-Ru
interaction where the Ru 4d electrons form itinerant-electron $\pi^{*}$ bands of t-orbital parentage. Presence of
competition between ferromagnetic and antiferromagnetic interactions is also evident from magnetization curves of
Co doped samples. Furthermore, the hysteresis loop recorded for 20\% Co doped compound at 5K in magnetic field
interval of $\pm$5 Tesla and presented in Fig. \ref{hys} exhibits a small irreversibility (see the expanded loop in the inset of Fig. \ref{hys}) indicating existence of
weak ferromagnetic interactions along with antiferromagnetic interaction. In the case of Ru doped compound
however, presence of clear magnetic hysteresis riding on smoothly increasing background indicates a strong
competition between ferro and antiferromagnetic interactions.

\begin{figure*}
\centering
\includegraphics[scale=0.75]{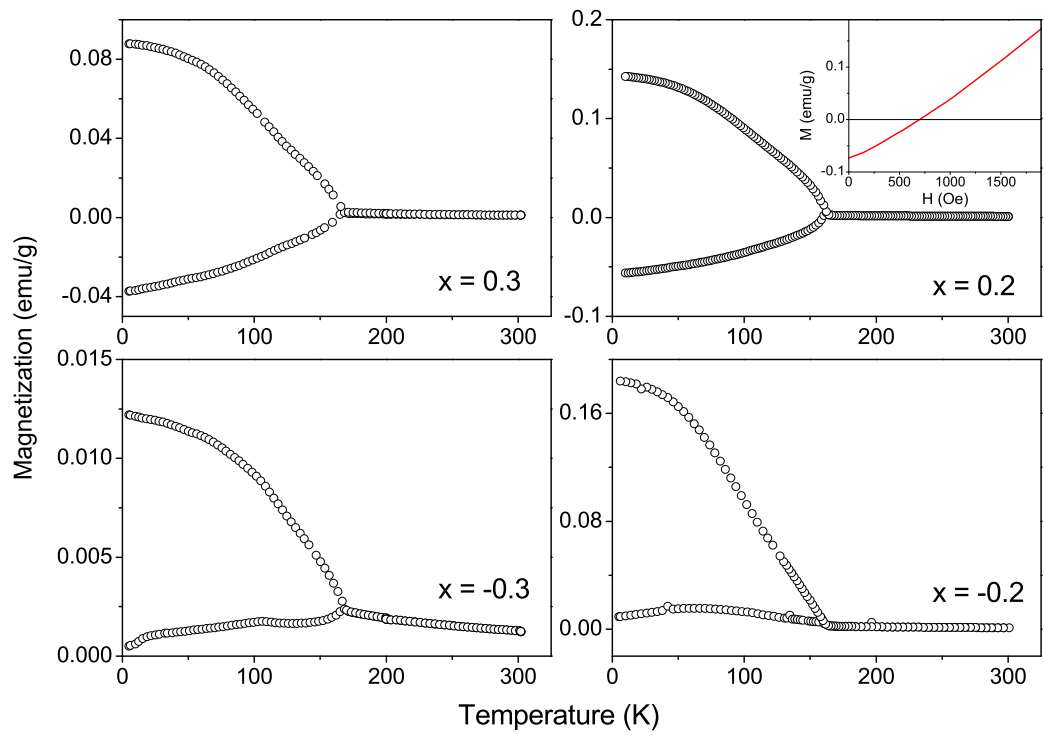}
\caption{\label{Mag(50Oe)} Low field (50 Oe) magnetization data as a function of temperature for LaSrCo$_{1-x}$Ru$_{1+x}$O$_6$.The inset in plot of $x$ = 0.2 shows the initial magnetization curve as a function of applied field for the same sample.}
\end{figure*}

In order to understand this competition further, magnetization was recorded in a low field of 50Oe. Here the
magnetization response in the Ru doped samples is completely different as compared to that measured at 1000 Oe as
can be seen in the Fig \ref{Mag(50Oe)}. During the ZFC cycle, the magnetization is negative at the lowest
temperature and increases continuously with increase in temperature. At about 158K and crosses over to the
positive side, exhibits a peak at 162K signifying a transition from a magnetically ordered to a paramagnetic
state. The magnetization remains positive during the entire FC cycle exhibiting a relatively sharp rise at 162K,
precisely the same temperature at which ZFC shows a cusp. In the case of Co doped samples, although there is a
large difference between ZFC and FC magnetization curves, the magnetization is positive during both cycles. It may be emphasized here that care has been taken to make sure that the remanent field of SQUID magnetometer was less than $\pm$13 Oe during these low field measurements. Furthermore, in the inset of M versus T plot for $x$ = 0.2 shows the initial magnetization curve for the same sample measured as a function of field. It can be clearly seen that magnetization  is negative in fields up to about 700 Oe.

\begin{figure*}
\centering
\includegraphics[scale=0.75]{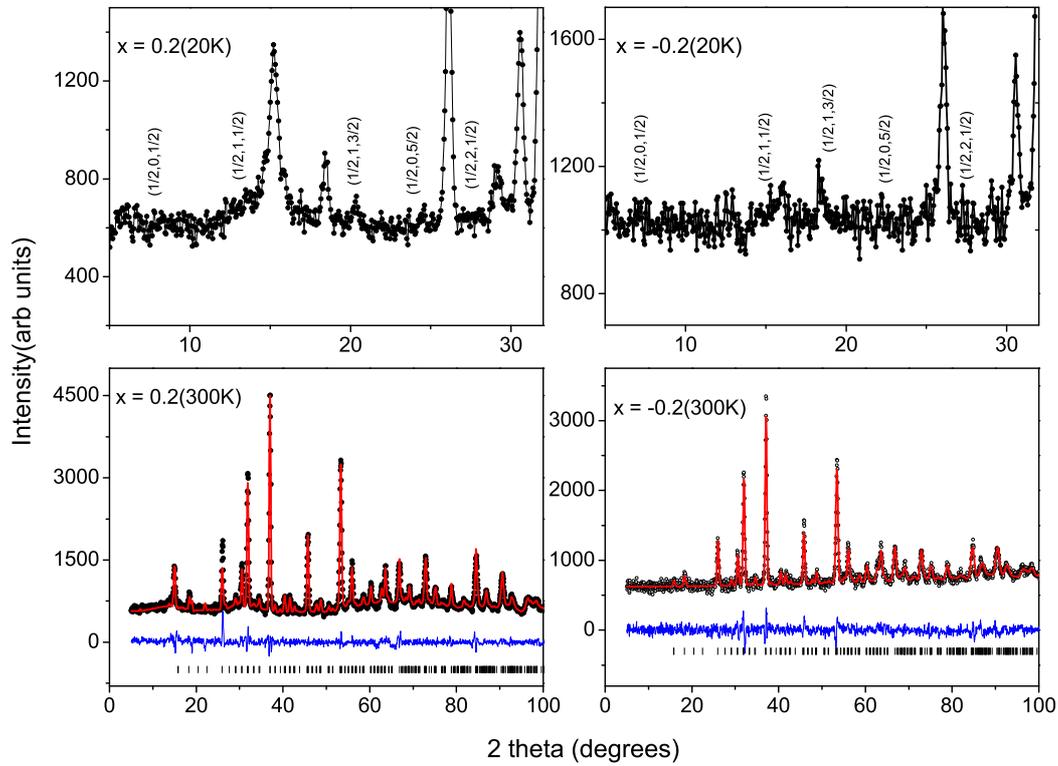}
\caption{\label{neutron} Neutron diffraction patterns recorded at 300K and 20K for $x$ = 0.2 and -0.2 of the series LaSrCo$_{1-x}$Ru$_{1+x}$O$_6$. The 300K data is presented in the 2$\theta$ range of 5$^\circ$ to 100$^\circ$ as circles along with Rietveld refined curve (solid line through the data) and the difference curve at the bottom. The 20K data is shown in limited range to highlight the weak magnetic reflections (marked with corresponding (h,k,l) values).}
\end{figure*}

Neutron diffraction was employed to determine the magnetic structure of two compounds, $x$ = 0.2 and -0.2. The
data taken at 300K and 20K is presented in Fig. \ref{neutron}. The data at 20K is shown in very limited range in
order to highlight the magnetic reflections. Very weak reflections due to AFM ordering were observed (marked in
Fig. \ref{neutron}) in the positions given by the propagation vector along the pseudocubic 111 direction with
respect to the crystallographic $P2_{1}/n$ cell \cite{bos,att}. The observation of the lines with k = (1/2, 0,
1/2) defines that AFM arrangement in each sublattice is of type II, consisting of ferromagnetic [111] planes. The values of magnetic moments at 20K obtained from neutron diffraction are $\mu(\rm{Co}) \sim \mu(\rm{Ru}) = 1.0 \mu_B$ for x = 0.2 and $\mu(\rm{Co}) \sim \mu(\rm{Ru}) = 0.6 \mu_B$ for x = -0.2.
This implies that the spins in (111) planes alternate as +/+/-/- for Co/Ru/Co/Ru. Such a AFM arrangement also
indicates that the magnetic ordering is governed by the 180$^\circ$ Co-O-Ru-O-Co AFM exchange paths
\cite{bosatt,rod}. Earlier studies report the presence of competing 90$^\circ$ Co-O-O-Co AFM exchange paths that
can be a cause of magnetic frustration leading to the low values of long-range-ordered moments as detected by
neutron scattering \cite{dlouha}.   

\begin{figure*}
\centering
\includegraphics[scale=0.5]{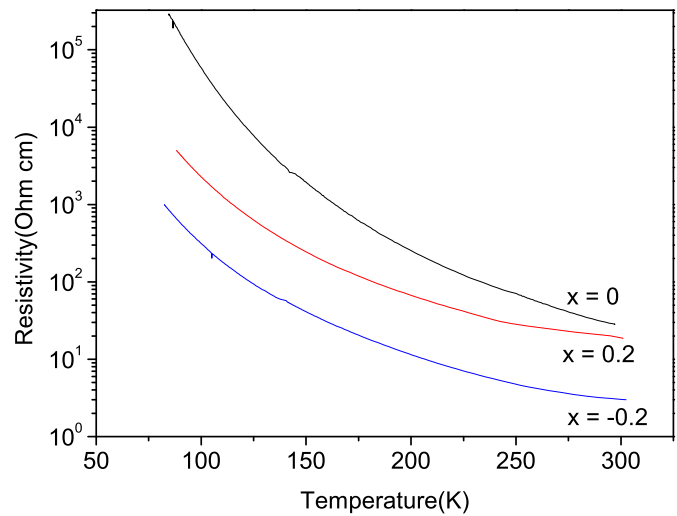}
\caption{\label{rest} Resistivity as a function of temperature for LaSrCo$_{1-x}$Ru$_{1+x}$O$_6$ compounds where $x$ = -0.2, 0 and 0.2}
\end{figure*}

The temperature dependence of the resistivity for  LaSrCo$_{1-x}$Ru$_{1+x}$O$_6$ with $x$ = -0.2, 0 and 0.2 are
plotted in the Fig \ref{rest}. The samples exhibit semiconducting behavior in the entire measured range of
temperature. The addition of Ru and Co dopants is found to decrease the resistivity significantly by about two
orders of magnitude.  In the case of Ru doped samples the decrease in resistivity is due to the formation of
metallic Ru$^{4+}$ - O - Ru$^{4+}$ chains in these samples which essentially result due to excess Ru
concentration. On the other hand in the case of Co doped samples, structural and magnetization studies indicated
the presence of Co$^{2+/3+}$ and Ru$^{4+/5+}$ redox pairs. The presence of Ru$^{4+}$ results in itinerant
electrons due to population of $\pi^*$ antibonding orbitals.  This explains the decrease in resistivity in case
of Co doped samples.

\subsection{LaSrCo$_{1-y}$Cu$_y$RuO$_6$}

\begin{figure*}
\centering
\includegraphics[scale=1]{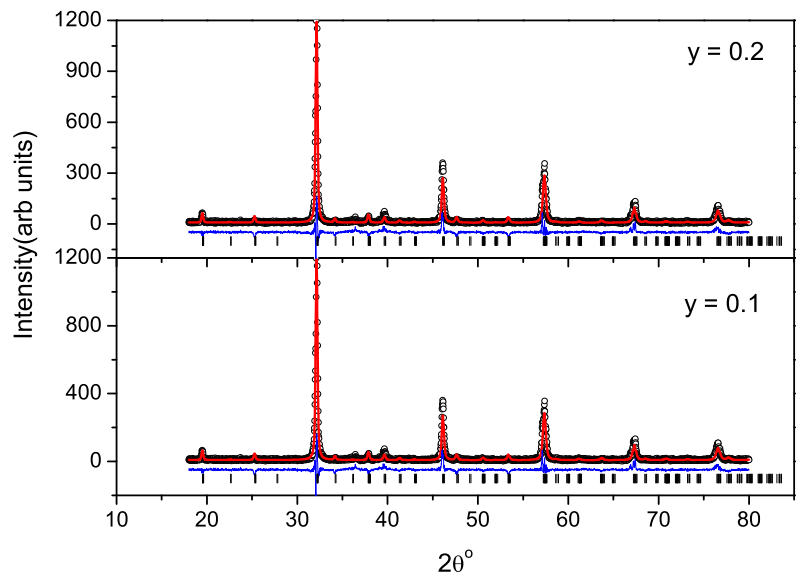}
\caption{\label{xrdCu} XRD patterns of LaSrCo$_{1-y}$Cu$_y$RuO$_6$ for $y$ = 0.1 and 0.2 along with best fitted line and difference line shown at the bottom. The ticks indicate the positions of allowed Bragg reflections.}
\end{figure*}

In Fig. \ref{xrdCu} XRD patterns of 10\% and 20\% Cu doped LaSrCoRuO$_6$ compounds is presented. Both these
compounds are single phase and the structural parameters obtained from Le Bail fitting are presented in Table
\ref{rietCu} along with that of undoped sample, LaSrCoRuO$_6$ for ready comparison. The unit cell volume in Cu
doped samples is slightly lower. This could be because of slightly smaller ionic radii of Cu$^{2+}$ (0.73\AA)
than that of Co$^{2+}$ (0.745\AA) in HS configuration. Substitution of Cu for Co is expected not to alter charge
balance and therefore Ru will be in 5+ state as in undoped compound. However, the substitution is expected to
alter magnetic and transport properties. Substitution of Cu in Sr$_2$HoRuO$_6$ type double perovskites is
reported to exhibit a coexistence of antiferromagnetism and superconductivity \cite{park}.

\begin{table}
\caption{\label{rietCu}Room temperature structural parameters obtained from Le Bail fitting and high temperature
magnetic and transport properties for LaSrCo$_{1-x}$Ru$_{1+x}$O$_6$. Here $a$, $b$, $c$ and $\beta$ denote unit
cell parameters, $\mu_{eff}$ and $\Theta_{CW}$ represent effective magnetic moment and Curie Weiss temperature
respectively and $T_p$ is temperature of peak magnetization seen in ZFC curve. Numbers in parentheses are
uncertainty in the last digit.}
\centering
\begin{tabular}{llll}
\hline\noalign{\smallskip}
$x$ & 0 & 0.1 & 0.2\\
\noalign{\smallskip}\hline\noalign{\smallskip}
$a$ (\AA)  & 5.5847(4)& 5.5621(2) & 5.5627(2)\\
$b$ (\AA) & 5.5592(6) & 5.5518(2) & 5.5544(2)\\
$c$ (\AA) & 7.8673(4) & 7.8684(2) & 7.8714(2)\\
$\beta$  & 90.05(2) & 90.23(1) & 90.28(1)\\
Volume (\AA$^3$) & 244.25(4) & 242.98(4) & 243.20(3)\\
$\mu_{eff}$ ($\mu_B$/fu) & 5.47(2) & 5.18(1) & 5.01(1)\\
$\Theta_{CW} (K)$ & -49(1) & -32.1(3) & -7.3(4)\\
$T_p$ (K) & 70(1) & 47(1) & 42(1) \\
\noalign{\smallskip}\hline
\end{tabular}
\end{table}

\begin{figure*}
\centering
\includegraphics[scale=0.75]{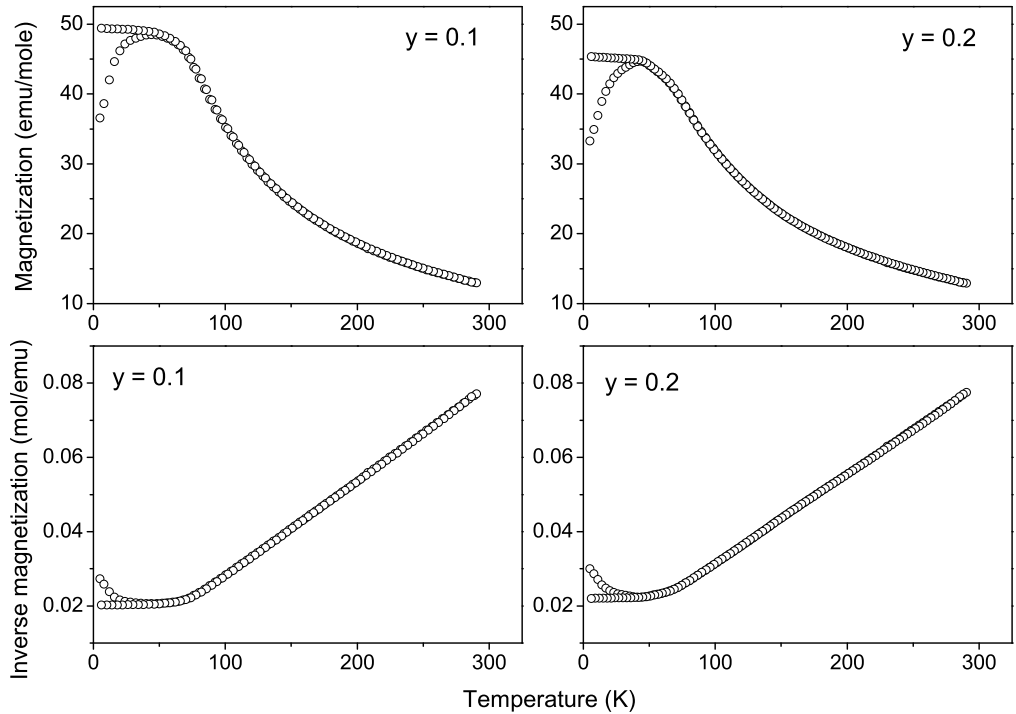}
\caption{\label{magCu} ZFC and FC magnetization curves for LaSrCo$_{1-y}$Cu$_y$RuO$_6$ recorded at H = 1000Oe.}
\end{figure*}

Magnetization for the two Cu doped samples, recorded during ZFC and FC cycles in an applied field of 1000 Oe is
presented in Fig. \ref{magCu}. In both cases both ZFC and FC magnetization curves increases with decreasing
temperature culminating into a broad peak centred at 42K and 47K respectively for $y$ = 0.1 and 0.2 samples.
Inverse magnetization plotted for both the samples in lower panel of Fig. \ref{magCu} shows linear variation down
to 100K. The calculated susceptibility ($\chi$ = M/H) was fitted to Curie-Weiss equation. The values of
$\mu_{eff}$ and $\Theta_{CW}$ extracted from fitting are listed in Table \ref{rietCu}. The values of
$\Theta_{CW}$ obtained are both negative although much smaller than that of the undoped sample. Negative values
indicate dominance of antiferromagnetic interactions in these compounds. Theoretically calculated spin only
moments using Ru$^{5+}$ (S = 3/2), Co$^{2+}$ (S = 3/2) and Cu$^{2+}$ (S = 1/2) for both the doped samples are in
good agreement with the experimentally obtained values. This behavior in unlike that seen in case of Ru and Cu
doped compounds. The large splitting observed between ZFC and FC magnetization curves is absent. Dominant
interaction is antiferromagnetic. Even with presence of a impurity ion (Cu$^{2+}$) there appears to be no
clustering or change in valence of Ru or Co.

\begin{figure*}
\centering
\includegraphics[scale=0.5]{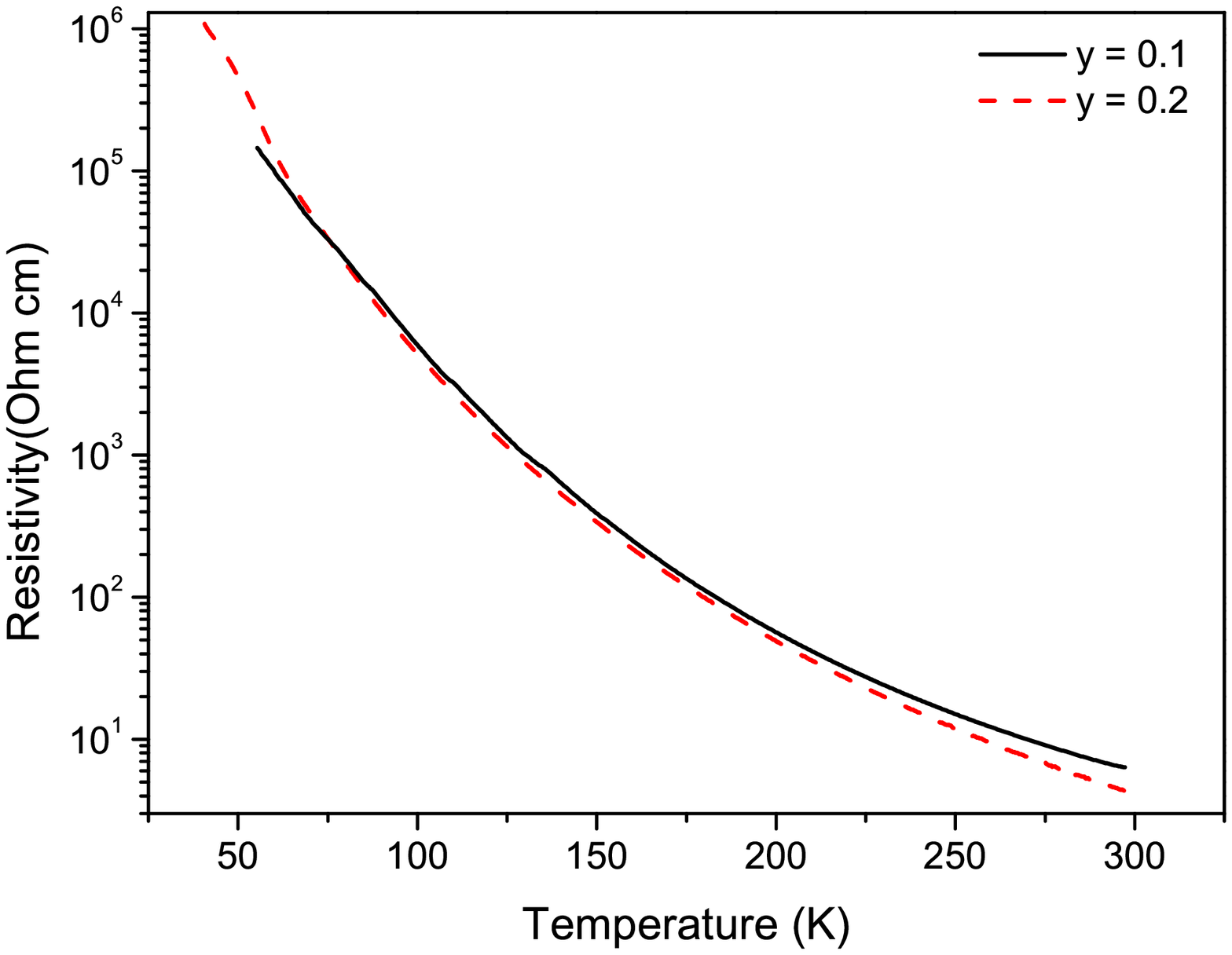}
\caption{\label{resCu}Temperature variation of electrical resistivity for LaSrCo$_{1-y}$Cu$_y$RuO$_6$ where $y$ =
0.1 and 0.2.}
\end{figure*}

In order to see the effect of Cu doping on the transport properties of LaSrCoRuO$_6$, electrical resistivity in
the temperature range 10K $<$T $<$ 300K has been measured and presented in Fig. \ref{resCu}. It appears that with
Cu substitution, there is very little change in resistivity as compared to the undoped sample. This can be
attributed to Cu$^{2+}$ replacing Co$^{2+}$ ions inhibiting the conversion of Ru$^{5+}$ to Ru$^{4+}$ or formation
of Ru-O-Ru chains.

\section{Discussion}
Substitution at B-site in ordered LaSrCoRuO$_6$ double perovskite allows an opportunity to probe the changes in
structural, magnetic and transport properties due to creation of Co$^{2+/3+}$ and Ru$^{4+/5+}$ redox couples in
an environment with fixed A-site variance and linearly varying Goldschmidt tolerance factor. In the entire doping
region, for both the series, the double perovskite structure is preserved.

The magnetic and transport properties of LaSrCo$_{1-x}$Ru$_{1+x}$O$_6$ on the other hand show significant changes
due to the creation of Ru$^{4+}$ and Co$^{3+}$ species in both Co doped and Ru doped compounds. The antisite disorder resulting due to excess Ru or Co leads to creation of Ru$^{4+}$ and Co$^{3+}$ species in addition to Ru$^{5+}$ and Co$^{2+}$ ions in order to maintain charge balance. In  the case of
undoped compound, LaSrCoRuO$_6$ there are two competing magnetic interactions, the linear Co-O-Ru-O-Co and the
90$^\circ$ Co-O-O-Co. In the Ru and Co doped compounds these interactions get diluted at the expense of new
interactions of the type Ru$^{4+/5+}$-O-Ru$^{4+/5+}$ and those involving HS/IS Co$^{3+}$. The sharp rise in
magnetization seen at about 160K in all Co and Ru doped compounds can be attributed to ferromagnetic interactions
arising due to Ru-O-Ru nearest neighbor interactions. As in case of SrRuO$_3$ such an interactions leads to
filling up of $\pi^*$ band and lowering of resistivity as compared to undoped LaSrCoRuO$_6$. The negative
magnetization displayed by Ru rich compounds in low applied fields can then be understood to be due to
polarization of paramagnetic Co spins by Ru-O-Ru ferromagnetic interactions below the Ru sublattice ordering
temperature (~160K). The Co spins are polarized in a direction opposite to applied magnetic field giving rise to
magnetic compensation and negative magnetization. In the case of Co doped compounds, no negative magnetization is
observed although there is the steep rise of magnetization below 160K. Instead, only a large variation between
ZFC and FC magnetization curves is seen. This could be understood to be due to clustering of similar types of
ions in the sample. This is also indicated by decrease in the strength of ferromagnetic interactions with
increasing Co content.

The competition between magnetic interactions observed in case of Co and Ru substituted compounds is reduced or eliminated with Cu substituting Co in LaSrCoRuO$_6$. The magnetic
properties clearly indicated presence of antiferromagnetic interactions from negative $\Theta_{CW}$ and the
magnetization behavior in the entire temperature range. Since Cu${2+}$ substitutes Co$^{2+}$, the charge on Ru
remains unaltered. Since the ionic sizes of the two ions are also similar, there are no major structural changes.
This results in nearly similar magnitude of resistivity as in the case of undoped LaSrCoRuO$_6$.

It is known that partially filled 3d levels of Co in SrRuO$_3$ are well below the conduction band of Ru$^{4+}$.
Therefore it is favorable to transfer a electron from Ru to Co, even though such electrons may become localized.
The resulting Co$^{2+}$ state will move its level closer to but still below Ru 4d levels. Thus the presence of
highly acidic Ru$^{5+}$ is more stabilizing for high spin Co$^{2+}$ leading to an ordered lattice. Doping of Co
and Ru will result in formation of Co-O-Co and Ru-O-Ru networks respectively. Form XANES studies reported in
\cite{mam} it is clear that Co-Co pairs will favor the trivalent state, and likewise Ru-Ru pairs would favor the
tetravalent states. Therefore the observed magnetic behavior reflects the general competition between the
itinerant ferromagnetism and the antiferromagnetic superexchange coupling between Co-Co pairs, further modified
by the interactions of between Ru and Co. Since Cu substitution does not favor formation of Ru-O-Ru or Co-O-Co
networks but instead disturbs the existing magnetic interactions especially the 90$^\circ$ Co-O-O-Co interaction,
antiferromagnetism appears to be strengthened.

\section{Conclusion}
The study of magnetic and transport properties of  ceramic samples LaSrCo$_{1-x}$Ru$_{1+x}$O$_6$ and
LaSrCo$_{1-y}$Cu$_y$RuO$_6$ leads to the following conclusions:
\begin{enumerate}
\item The ordered double perovskite phase exists for $-0.3 \le x \le 0.4$.
\item The thermally activated electrical resistivity is associated with nearest neighbor hopping between Co and
Ru ions via linking oxygen atoms. The decrease in resistivity with Ru and Co doping is due to formation of
Ru$^{4+}$ and more holes in the Ru $d$ band.
\item The Co and Ru doped compounds also show an increase in magnetization due to the formation of ferromagnetic Ru-O-Ru
interactions.
\item In case of Cu doped compounds, the antiferromagnetism strengthens due to the suppression of the
competing 90$^\circ$ Co-O-O-Co antiferromagnetic interactions.
\end{enumerate}

\begin{acknowledgement}
PSRM and KRP acknowledge support from UGC-DAE Consortium for Scientific Research, Mumbai Centre for financial
support under CRS-M-126. KRP and PRS would also like to thank Department of Science and Technology (DST),
Government of India for financial support under the project No. SR/S2/CMP-42.
\end{acknowledgement}

\begin{thebibliography}{100}
\bibitem{dd} D. D. Sarma, Curr. Opin. Solid State Mater. Sci \textbf{5} (2001) 261
\bibitem{koba} K. L. Kobayashi, T. Kimura, H. Sawada, K. Terakura, Y. Tokura Nature \textbf{395} (1998) 677
\bibitem{jg} J. Gopalakrishnan, A. Chattopadhyay, S. B. Ogale, T. Venkatesan, R. L. Greene, A. J. Millis, K. Ramesha, B. Hannoyer and G. Marest, Phys. Rev. B \textbf{62} (2000) 9538
\bibitem{sriti} F. Sriti, A. Maignan, C.  Martin and B. Raveau, Chem. Mater. {\bf 13} (2001) 1746
\bibitem{feng} X. M. Feng, G. H. Rao, G. Y. Liu, H. F. Yang, W. F. Liu, Z. W. Ouyang, L. T. Yang, Z. X. Liu, R. C. Yu, C. Q. Jin and J. K. Liang, J. Phys. Condens. Matter {\bf 14} (2002) 12503
\bibitem{battle} P. D. Battle and W. J. Macklin, J. Solid State Chem. {\bf 52} (1984) 138
\bibitem{kame} N. Kamegashira, T. Mori, A. Imamura, and Y. Hinatsu, J. Alloys Compd. {\bf 302}, (2002) L6
\bibitem{doi} Y. Doi, Y. Hinatsu, K. Oikawa, Y. Shimojo and Y. Morii, J. Mater. Chem. {\bf 10} (2000) 797
\bibitem{wu} M. K. Wu, D. Y. Chen, F. Z. Chien, S. R. Sheen, D. C. Ling, C. Y. Tai, G. Y.  Tseng, D. H. Chen and F. C. Zhang, Z. Phys. B: Condens. Matter {\bf 102} (1997) 37
\bibitem{park} N. G. Parkinson, P. D. Hatton, J. A. K. Howard, C. Ritter, F. Z. Chien and M. K. Wu, J. Mater. Chem. {\bf 13} (2003) 1468
\bibitem{dass} R. I. Dass, J. Q. Yan and J. B. Goodenough, Phys. Rev. B {\bf 69} (2004) 094416
\bibitem{bos} J-W. G. Bos and J. P. Attfield, Chem. Matter. {\bf 16} (2004) 1822
\bibitem{mam} A. Mamchik, W. Dmowski, T. Egami and I-W Chen, Phys. Rev. B {\bf 70} (2004) 104410
\bibitem{tomes} P. Tome$\check{\rm s}$, J. Hejtm$\acute{\rm a}$nek and K. Kn$\acute{\rm i}\check{\rm z}$ek,  Solid State Sci. {\bf 10} (2008)486
\bibitem{dlouha} M. Dlouh$\acute{\rm a}$, J. Hejtm$\acute{\rm{a}}$nek, Z. Jir$\acute{\rm{a}}$k, K. Kn$\acute{\rm{i}}$$\check{\rm{z}}$ek, P. Tome$\check{\rm s}$ and S. Vratislav S J. Magn. Magn. Mater. (2009) doi:10.1016/j.jmmm.2009.06.035
\bibitem{serrate} D. Serrate, J. M. de Teresa and M. R. Ibarra, J. Phys. Condens. Matter {\bf 19} (2007) 023201
\bibitem{gold} V. M. Goldschmidt, Skrifter Nordske Videnskaps-Akad. Oslo I, Mat-Naturvidensk Kl., {\bf 8} (1926) 2
\bibitem{kim} S. H. Kim and P. D. Battle, J. Sol. State. Chem {\bf 114} (1995) 174
\bibitem{att} J-W. G. Boss and J. P. Attfield, J. Mater. Chem. {\bf 15} (2005) 715
\bibitem{bosatt} J-W. G. Bos, and J. P. Attfield, Phys. Rev. B \textbf{70} (2004) 174434
\bibitem{rod} E. Rodriguez, M. L. Lopez, J. Campo, M. L. Veiga and C. Pico, J. Mater. Chem. \textbf{12} (2002) 2798
\end{thebibliography}
\end{document}